\documentclass[11pt,twoside]{article}

\usepackage{asp2006}
\usepackage{epsf}
\usepackage{psfig}
\usepackage{lscape}

\newcommand{\ergs}{ergs s$^{-1}$}
\newcommand{\flux}{ergs cm$^{-2}$ s$^{-1}$}

\newcommand{\cdens}{cm$^{-2}$}

\newcommand{\chandra}{{\it Chandra}}

\newcommand{\bootes}{Bo\"{o}tes}
\newcommand{\spitzer}{{\it Spitzer}}

\begin{document}
\title{A rich bounty of AGN in the 9 deg$^2$ \bootes\ survey: high-$z$
  obscured AGN and large-scale structure}   
\author{Ryan C. Hickox\altaffilmark{1}, Christine Jones\altaffilmark{1}, William R. Forman\altaffilmark{1}, Stephen
  S. Murray\altaffilmark{1}, Almus Kenter\altaffilmark{1}, Mark
  Brodwin\altaffilmark{2}, and the \chandra\ X\bootes, NOAO Deep Wide-Field Survey, {\it Spitzer} IRAC Shallow
  Survey, and AGES Teams}   
\altaffiltext{1}{Harvard-Smithsonian Center for Astrophysics, 60 Garden
  Street, Cambridge, MA  02138, USA}   
\altaffiltext{2}{Jet Propulsion Laboratory, California Institute of
  Technology, Pasadena, CA 91109, USA}   

\begin{abstract} 
We use observations from the 9 square degree multiwavelength survey in
\bootes\ to identify hundreds of obscured active galactic nuclei
(AGN) with high redshifts ($z>0.7$), luminosities ($L_{\rm bol}>10^{45}$ \ergs), and
moderate obscuring columns ($N_{\rm H}> 10^{22}$ \cdens), and to
measure the clustering properties of X-ray AGN at $z>1$.    In the
\bootes\ region, shallow (5 ks) \chandra\ X-ray observations have
detected $\sim$4,000 X-ray sources, and the same region
has been mapped with deep optical imaging and by \spitzer\ IRAC, which
detects $\sim$300,000 point sources, of which $\sim$30,000 have detections in all four IRAC bands, for which we can select
AGN on the basis of their mid-IR colors.  With the MMT/Hectospec
we have obtained modest resolution optical spectra for
about half the X-ray sources (out to $z>3$) and $\sim$20,000 galaxies
(out to $z=0.7$).  With this multiwavelength data we select $>$400 AGN
per square degree (compared to 12 per square degree from SDSS).  Among
a sample of IRAC-selected AGN we identify 641 candidate obscured
objects based on their $R$ band and IRAC luminosities. We use X-ray
stacking techniques to verify that they are obscured AGN and measure
their absorbing column densities.  We also measure the
three-dimensional two-point correlation function for X-ray selected
AGN.

\end{abstract}



\section{The \bootes\ multiwavelength dataset}
In studies of AGN, wide-field surveys are complementary to deep
surveys such as the GOODS fields in that they explore the rare, bright
end of the luminosity function and constrain spatial distributions
over large angular scales.  In the \bootes\ region, shallow (5 ks)
\chandra\ X-ray observations have detected $\sim$4,000 X-ray sources, of
which most are AGN, to a limiting 0.5--7 keV flux of about
$4\times10^{-15}$ \flux\ \citep{murr05, kent05}.  The region has also
been mapped by deep $B_W$, $R$, and $I$ imaging with the NOAO Deep
Wide-Field Survey \citep[NDWFS,][]{jann99}, and by the \spitzer\ IRAC
Shallow Survey, which detects $\sim$300,000
point sources, of which $\sim$30,000 have 5$\sigma$ detections in all
four IRAC bands, and for which AGN can be selected on the basis of
their mid-IR colors \citep{eise04}.  Using the MMT Hectospec
multi-fiber spectrograph, the AGES survey has obtained modest resolution
optical spectra for about half the X-ray sources out to $z>3$ and
$\sim$20,000 galaxies out to z=0.7 (Kochanek et~al. 2006, in
preparation).  These optical, infrared, and X-ray techniques select
$>$400 AGN per square degree.

Redshifts for objects are obtained spectroscopically using the AGES
spectra, or using photometric redshift estimators based on the IRAC
and optical photometry \citep{brod06}.
Photo-$z$'s have uncertainties $\sigma_{z}=0.06(1+z)$ for galaxies
at $z<1$, and $\sigma_{z}=0.12(1+z)$ for optically bright AGN.

\section{Identifying obscured AGN}
In the standard picture of AGN emission, observations at X-ray,
optical, and IR wavelengths correspond to different regions in the
central engine.  X-rays are emitted in the inner regions close to the
black hole, the optical/UV continuum is emitting by the accretion
disk, and optical broad and narrow lines are produced by illuminated
gas surrounding the accretion disk.  Infrared emission is produced by
the reprocessing of UV and X-ray emission by surrounding dust.

\begin{figure}
\plottwo{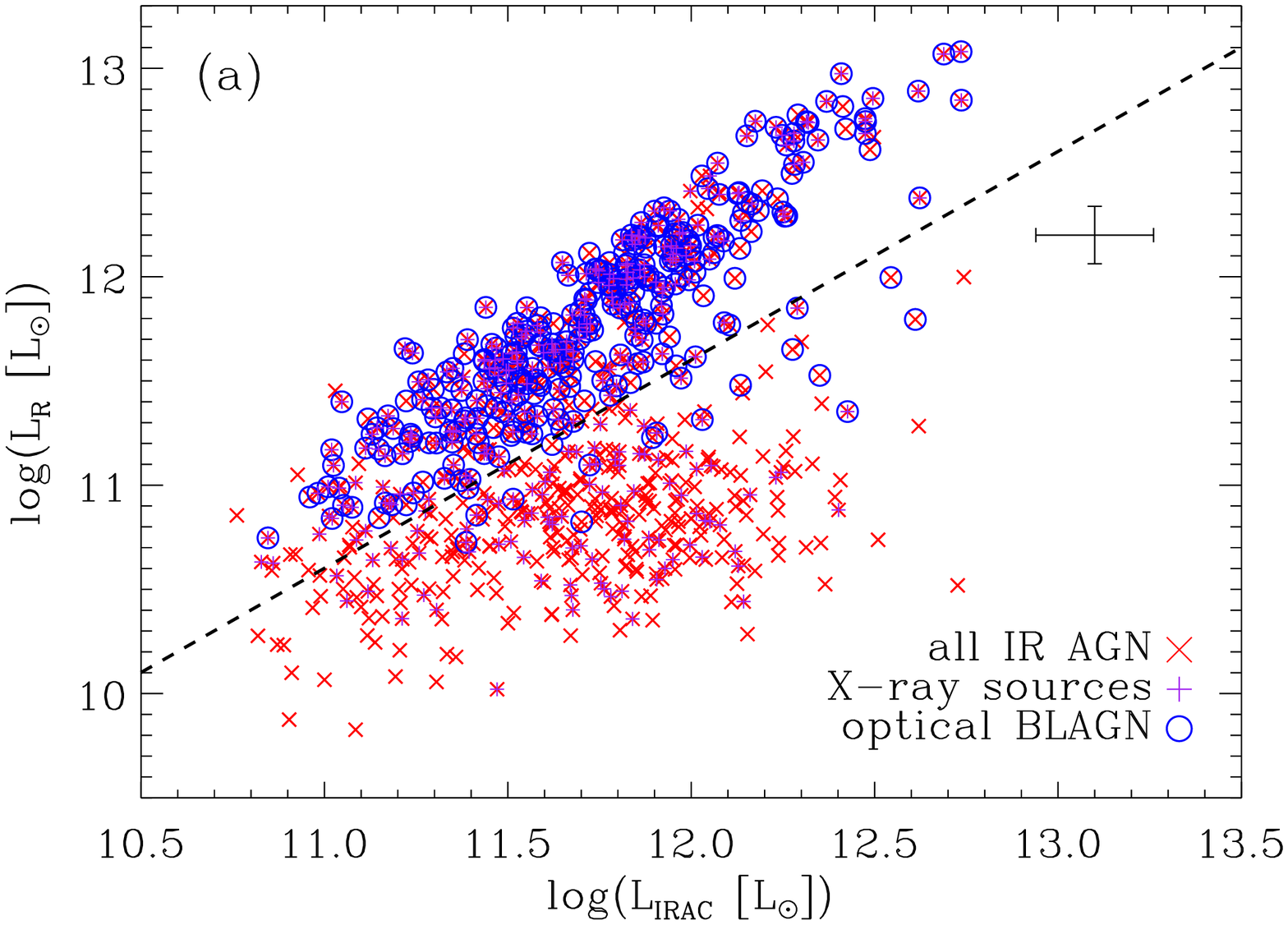}{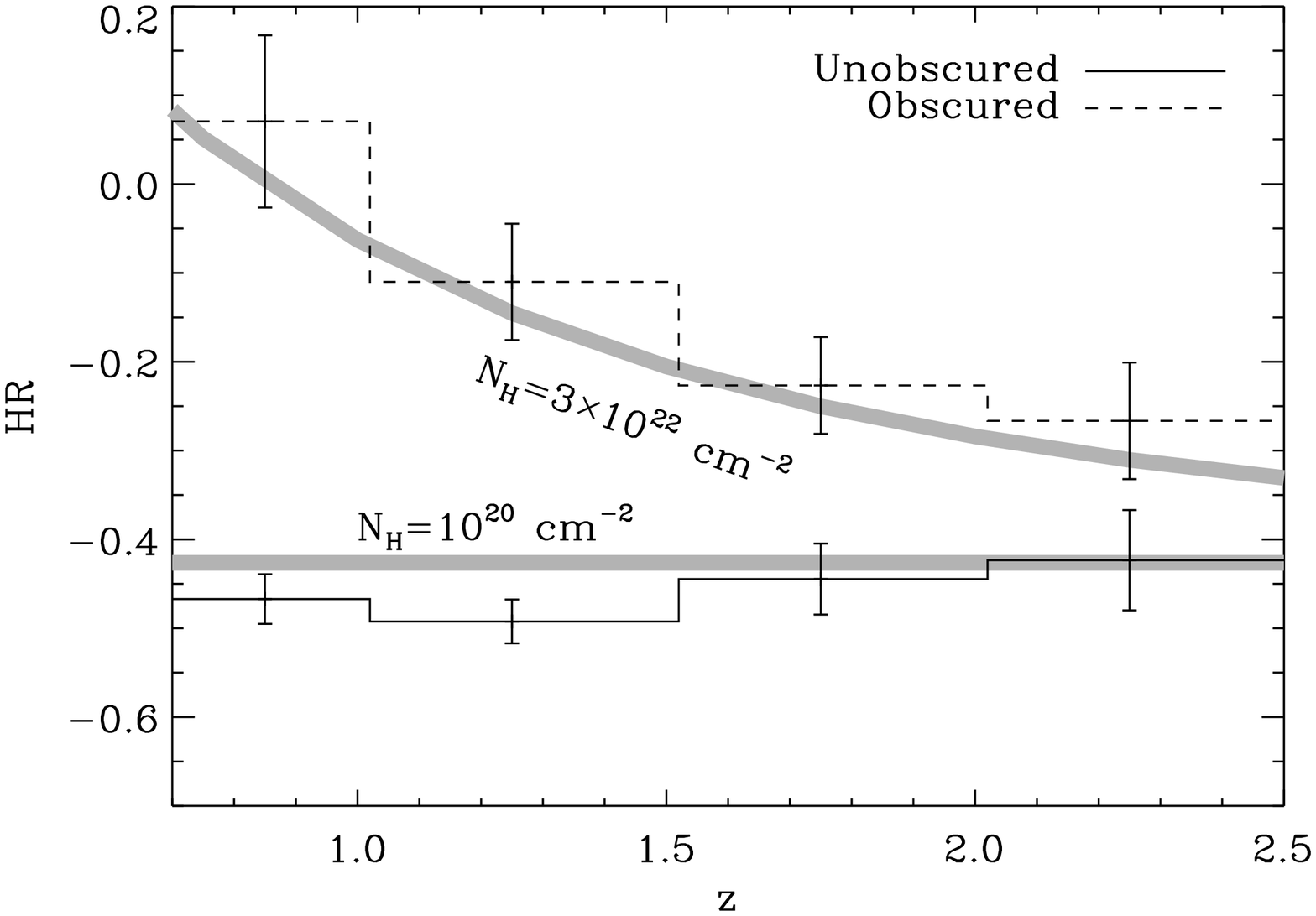}
\caption{ {\it Left:} $R$ band versus mid-IR luminosity for
  1410 infrared-selected AGN from the \bootes\ field.  The dashed line
  shows an empirical criterion selecting obscured
  (bottom) and unobscured (top) AGN. {\it Right:} Hardness ratio
  versus $z$ for the two types of IR-selected AGN.  Gray lines
  show $HR$ for an intrinsic $\Gamma=1.8$ and
  two values of $N_{\rm H}$.  The obscured objects have
  systematically harder X-ray spectra, corresponding to $N_{\rm
  H}\sim3\times10^{22}$ cm$^{-2}$.  \label{figiagn}}
\end{figure}

Methods for detecting AGN on the basis of their optical and X-ray
emission are well established.  Recently, new techniques using IRAC
have allowed the identification of AGN on the basis of their observed
colors in the mid-infrared, where AGN tend to have a red power-law
continuum that separates them from starburst galaxies in color-color
space \citep[e.g.,][]{lacy04, ster05}.  Because mid-IR emission is
less affected by obscuration from dust and gas than the optical or
soft X-rays, this technique allows the detection of obscured AGN that may
be missed with other techniques \citep[e.g.,][]{alon06, mart06}.

We identify obscured AGN from a color-color selected sample using
observed optical and IR luminosities \citep{hick06c}.  Fig.\
\ref{figiagn} shows the distribution in observed $\nu L_{\nu}$ in the
$R$ band and IRAC bands for 1410 IRAC-selected AGN at $z>0.7$.  The
luminosity distribution shows two clear populations: (1) objects in
which $L_R$ increases with $L_{\rm IRAC}$, and are associated with
sources having optical broad-line AGN spectra, and (2)  objects with
roughly constant $L_R$ over two orders of magnitude in $L_{\rm IRAC}$.
We classify the first population of 769 objects as unobscured AGN
(IRAGN 1) and the second 641 objects as AGN for which the optical
light is obscured (IRAGN 2), with a boundary between them of
$\log{L_R/L_{\rm IRAC}}=-0.4$.

We verify this simple selection of obscured AGN using (1) the
\chandra\ X-ray observations, and (2) the observed optical colors and
morphologies.  Because the \chandra\ observations are shallow, most
objects have no more than a few X-ray counts.  Therefore, we perform
stacking analyses to determine the average fluxes and spectra of the
different subsets of objects.  We find that the IRAGN 1s and IRAGN 2s
have similar X-ray luminosities ($L_X\sim10^{43}$--$10^{44}$ \ergs),
but the IRAGN 2s have significantly harder X-ray spectra, shown by a
larger value of the hardness ratio $HR=(H-S)/(H+S)$, where $H$ and $S$
are observed counts in the 0.5--2 and 2--7 keV bands, respectively
(Fig.\ \ref{figiagn}).  Assuming a typical AGN power law X-ray spectrum with
photon index $\Gamma=1.8$, this indicates absorption by a neutral
hydrogen column of $N_{\rm H}\sim3\times10^{22}$ \cdens, which is
typically in the range of ``obscured'' AGN in population models \citep[e.g.,][]{trei05}.

\begin{figure}
\psfig{figure=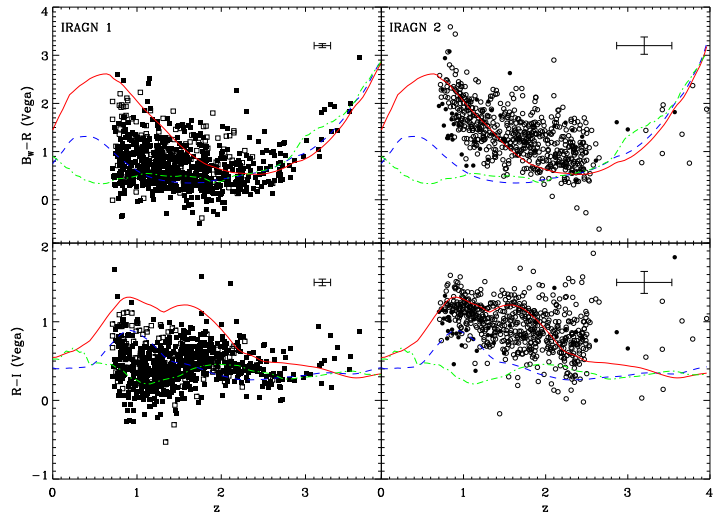,width=10cm}
\caption{Observed $B_{W}-R$ and $R-I$ colors for IRAGN 1s (left) and
  IRAGN 2s (right), compared to elliptical galaxy (red solid line), Sb
  galaxy (blue dashed line),
  and quasar (green dot-dashed line) templates.  Filled symbols show objects
  with optical spectroscopic redshifts, while empty symbols have only
  photo-$z$s. \label{figcol}}
\end{figure}

The optical colors of the IRAGN 2s are also typical of normal galaxies
rather than nuclear AGN emission, indicating that the nuclear optical
light is obscured.  Fig.\ \ref{figcol} shows 
$B_W-R$ and $R-I$ color versus redshift for the two IRAGN types
compared with quasar \citep{vand01} and galaxy \citep{fioc97} templates.  The
IRAGN 2s are redder than would be expected for an unobscured quasar.
Also, the IRAGN 2s have extended morphologies,
while the unobscured IRAGN 1s are dominated by point-like nuclear
emission.

The X-ray and optical properties of the IRAGN 2 sample therefore are
consistent with these 641 objects being a population of AGN with high
luminosities ($L_{\rm bol}>10^{45}$ \ergs) redshifts ($z>0.7$), and
moderate obscuration ($N_{\rm H}>10^{22}$ \cdens).  This one of the
largest such samples identified to date, made possible by the wide
field and extensive multiwavelength data in the \bootes\ region.

\section{AGN clustering properties}
The contiguous X-ray coverage in the \bootes\ field has allowed us to
spectroscopically target, and obtain accurate redshifts, for a large
number of AGN at $z>0.7$, and so observe large-scale structure out to
$z=1$ and beyond.  We have determined the three-dimensional two-point
correlation function for the X-ray selected AGN, and find significant
evidence for clustering (Fig.\ 3, Kenter et~al. 2006, in preparation).  The inferred clustering
parameters are consistent with those found by the optically-selected
2dF Quasar survey \citep{croo05}.


\begin{figure}

\psfig{figure=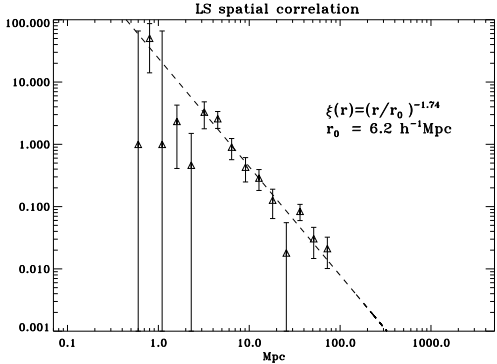,width=7cm}
\caption{Three-dimensional two-point correlation function for X-ray selected
AGN from the X\bootes\ survey.  Redshifts are obtained from optical
spectroscopy with the AGES survey. \label{figstruct}}
\end{figure}

\acknowledgements 
We thank all the members of the NOAO DWFS, X\bootes, IRAC Shallow
Survey, and AGES teams.  RCH is supported by a NASA GSRP Fellowship
and a Harvard Merit Fellowship.  



\end{document}